# Cooling mechanisms in molecular conduction junctions

Michael Galperin, Department of Chemistry & Biochemistry, University of California at San Diego, La Jolla, CA, 92093.
Keiji Saito, Graduate School of Science, University of Tokyo, 113-0033, Japan, and CREST, Japan Science and Technology (JST), Saitama, 332-0012, Japan.
Alexander V. Balatsky, Theoretical Division, Los Alamos Natinal Laboratory, Los Alamos, NM, 87545.
Abraham Nitzan, School of Chemistry, Tel Aviv University, Tel Aviv, 69978, Israel

## Absract

While heating of a current carrying Ohmic conductors is an obvious consequence of the diffusive nature of the conduction in such systems, current induced cooling has been recently reported in some molecular conduction junctions. In this paper we demonstrate by simple models the possibility of cooling molecular junctions under applied bias, and discuss several mechanisms for such an effect. Our model is characterized by single electron tunneling between electrodes represented by free electron reservoirs through a system characterized by it electron levels, nuclear vibrations and their structures. We consider cooling mechasims resulting from (a) cooling of one electrode surface by tunneling induced depletion of high energy electrons; (b) cooling by coherent sub resonance electronic transport analogous to atomic laser nduced cooling and (c) the incoherent analog of process (b) – cooling by driven activated transport. The non-equilibrium Green function formulation of junction transport is used in the first two cases, while a master equation approach is applied in the analysis of the third.

# 1 Introduction

One of the main problems of practical molecular electronics is junction heating caused by vibrational excitation by the electron flux. Consequently, heat transport in mesoscopic systems attracts considerable attention both experimentally[1-5] and theoretically.[6-33] Upon increasing the voltage bias $V$, junction heating starts above the threshold, $eV = \hbar\omega$ for vibrational excitation and increases further beyond the conduction threshold where a bridge level enters the Fermi window. If energy dissipation is not effective enough, junction temperature increases and its stability may be jeopardised. Therefore heat transport away from the junction must be made as efficient as possible.

While heating in current carrying junctions seems natural, it is not a fundamental consequence of charge transport. Here we focus on the relationship between applied bias, current flow and temperature change. A similar question was recently addressed by Pistolesi,[34] who, however, imposed at the outset a lower than ambient electronic temperature thus rendering the issue somewhat trivial. Here we focus on the possibility that molecular vibrations may cool down by the electronic flux, where without voltage bias the system is at thermal equilibrium. The effect of junction cooling was considered so far mostly for superconductor-insulator-superconductor (SIS) and superconductor-insulator-normal metal (SIN) junctions.[35-39] The possibility of cooling in a metallic nanowire connected to superconductor leads due to coupling to longitudinal phonon modes was recently discussed.[40] In another recent work[41] Raman scattering was used to monitor molecular temperature in current carrying junctions, showing apparent cooling in certain bias regimes. Finally, Zippilli et al[42] have described a theoretical model for current induced cooling in carbon nanotubes.

Here we consider simple models of single-molecule conduction junctions, where a molecule connects between two metallic or semiconductor electrodes. Our goal is to find conditions under which molecular cooling under applied bias is possible. Our model systems are described next. Using these models we consider several possibilities for current induced molecular cooling.

## 2. Model

The models considered in this paper are special cases of a general molecular junction model that comprises a left ($L$) and right ($R$) electrodes represented as free electron reservoirs bridged by a molecule ($M$). The corresponding Hamiltonian is

$$\hat{H} = \hat{H}_L + \hat{H}_M + \hat{H}_R + \hat{H}_T + \hat{H}_B \tag{1}$$

where

$$\hat{H}_K = \sum_{k \in K} \varepsilon_k \hat{c}_k^\dagger \hat{c}_k \quad (K = L, R) \tag{2}$$

are the hamiltonians of the two electrodes and

$$\hat{H}_M = \sum_m \varepsilon_m \hat{d}_m^\dagger \hat{d}_m + \sum_\alpha \omega_\alpha \hat{a}_\alpha^\dagger \hat{a}_\alpha + \sum_{m,m',\alpha} M_{mm'}^\alpha \hat{Q}_\alpha \hat{d}_m^\dagger \hat{d}_{m'} \tag{3a}$$

is the molecular Hamiltonian. Here $\hat{c}_k^\dagger$ ($\hat{c}_k$) and $\hat{d}_m^\dagger$ ($\hat{d}_m$) are respectively creation (annihilation) operators of electrons in the metal states $k$ and in the molecular orbital $m$ with single electron energies $\varepsilon_k$ nad $\varepsilon_m$, $\hat{a}_\alpha^\dagger$ and $\hat{a}_\alpha$ are similar operators for the molecular vibrational normal mode $\alpha$ of frequency $\omega_\alpha$, and $\hat{Q}_\alpha = \hat{a}_\alpha + \hat{a}_\alpha^\dagger$ are the corresponding displacement operators. The last term in Eq. 3 is the electron-vibration coupling on the molecule. Note that molecular Hamiltonian (3a) is written in the representation of the eigenstates of the free molecule. Alternatively, local state represenation is often preferred, whereupon

$$\hat{H}_M = \sum_m \varepsilon_m \hat{d}_m^\dagger \hat{d}_m + \sum_{\substack{m,m' \\ m \neq m'}} V_{mm'} \hat{d}_m^\dagger \hat{d}_{m'} + \sum_\alpha \omega_\alpha \hat{a}_\alpha^\dagger \hat{a}_\alpha + \sum_{m,m',\alpha} M_{mm'}^\alpha \hat{Q}_\alpha \hat{d}_m^\dagger \hat{d}_{m'} \tag{3b}$$

The molecule is coupled to the electrodes $L$ and $R$ by an electron transfer Hamiltonian $\hat{H}_T$. Most generally it has the form

$$\hat{H}_T = \sum_{K=L,R} \sum_{k \in K} \sum_{m \in M} \left( V_{km} \hat{c}_k^\dagger \hat{d}_m + V_{mk} \hat{d}_m^\dagger \hat{c}_k \right) \tag{4}$$

We shall sometime consider the simplest model with only one relevant molecular electronic orbital $\varepsilon_0$ and one molecular vibration $\omega_0$. In this case

$$\hat{H}_M = \varepsilon_0 \hat{d}^\dagger \hat{d} + \omega_0 \hat{a}^\dagger \hat{a} + M \hat{Q}_0 \hat{d}^\dagger \hat{d} \tag{5}$$

$$\hat{H}_T = \sum_{K=L,R} \sum_{k \in K} \left( V_k^* \hat{c}_k^\dagger \hat{d} + V_k \hat{d}^\dagger \hat{c}_k \right) \tag{6}$$

Finally, the molecular vibrations may be coupled to a thermal bath, represented by

a set of harmonic oscillators $\{\omega_\beta\}$

$$\hat{H}_B = \sum_\beta \left( \omega_\beta \hat{b}_\beta^\dagger \hat{b}_\beta + \sum_\alpha U_{\alpha\beta} \hat{Q}_\alpha \hat{Q}_\beta \right) \tag{7}$$

where $\hat{Q}_\beta = \hat{b}_\beta + \hat{b}_\beta^\dagger$ are displacement operators for the bath vibrations.

## 3. Metal-insulator-metal junction

Consider first a junction in which the two metallic electrodes are coupled to one another directly (tunneling through an inert insulator). In such junction effective cooling of one (say, $L$) side and simultaneous heating of the other ($R$) side is possible if the electron transmission probability is higher for higher energy electrons. Such behavior is characteristic of tunneling through the barrier,[43-45] and can be caused e.g. by presence of resonant state within the insulator. Then at steady state, in proximity of the junction, a nonequilibrium electronic distribution will be established in the metallic contacts. For simplicity we model this behavior by splitting the electrodes Hamiltonians $H_K$ ($K = L, R$) into two parts

$$\hat{H}_K = \hat{H}_K^{eq} + \hat{H}_K^{neq} + \hat{V} \tag{8}$$

$$\hat{H}_K^q = \sum_{k \in K_q} \varepsilon_k \hat{c}_k^\dagger \hat{c}_k \; ; \; q = eq, neq \tag{9}$$

$$\hat{V} = \sum_{k \in K_{eq}} \sum_{k' \in K_{neq}} \left( V_{kk'} \hat{c}_k^\dagger \hat{c}_{k'} + \text{h.c.} \right) \tag{10}$$

where the equilibrium part describes the bulk of the electrodes, while non-equilibrium may prevail near the electrode surface, where depletion of electrons takes place by tunneling through the molecular layer. Thus the junction is modeled by sequence $L_{eq} - L_{neq} - R_{neq} - R_{eq}$, where both equilibrium and non-equilibrium parts of each electrode are assumed to be free electron reservoirs. The Fermi distribution

$$f_K^{eq}(E) = \left[ \exp\left( \frac{E - \mu_K}{k_B T} \right) + 1 \right]^{-1} \tag{11}$$

define the equilibrium states of the electrodes $K = L, R$, while the non-equilibrium segments near the junction are characterized by some nonequilibrium distributions

$f_K^{neq}(E)$ that need to be determined. To determine these non-equilibrium electronic distributions we use a system of rate equations for the energy-resolved populations $f_K^{neq}(E,t)$ ($K = L, R$), and consider a steady-state situation (see Appendix A for derivation)

$$\frac{dN_L^{neq}(E,t)}{dt} = \frac{T_{LL}}{\hbar}\left[f_L^{eq}(E) - f_L^{neq}(E)\right] + \frac{T_{LR}(E)}{\hbar}\left[f_R^{neq}(E) - f_L^{neq}(E)\right] = 0$$
$$\frac{dN_R^{neq}(E,t)}{dt} = \frac{T_{RR}}{\hbar}\left[f_R^{eq}(E) - f_R^{neq}(E)\right] + \frac{T_{LR}(E)}{\hbar}\left[f_L^{neq}(E) - f_R^{neq}(E)\right] = 0$$
(12)

where $N_K^{neq}(E,t) \equiv f_K^{neq}(E,t) A_K(E)$ and $A_K(E)$ is the density of states in the leads $K=L,R$. The (dimensionless) rate constants $T_{LL}$, $T_{RR}$ between the equilibrium and nonequilibrium parts at each electrode and $T_{LR}$ between the nonequilibrium segments on opposite electrodes are defined in Appendix A. We use the wide band approximation to describe the electronic continua in the leads. Inelastic processes are disregarded or their effects are assumed to be embedded into the rates $T_{LL}$ and $T_{RR}$. These rates are assumed to be energy independent, while the energy dependence of $T_{LR}(E)$ is taken to reflect tunneling through a square barrier

$$T_{LR}(E) = T_{LR}^0 \exp\left[-d\sqrt{\frac{2m}{\hbar^2}(U-E)}\right] \qquad (13)$$

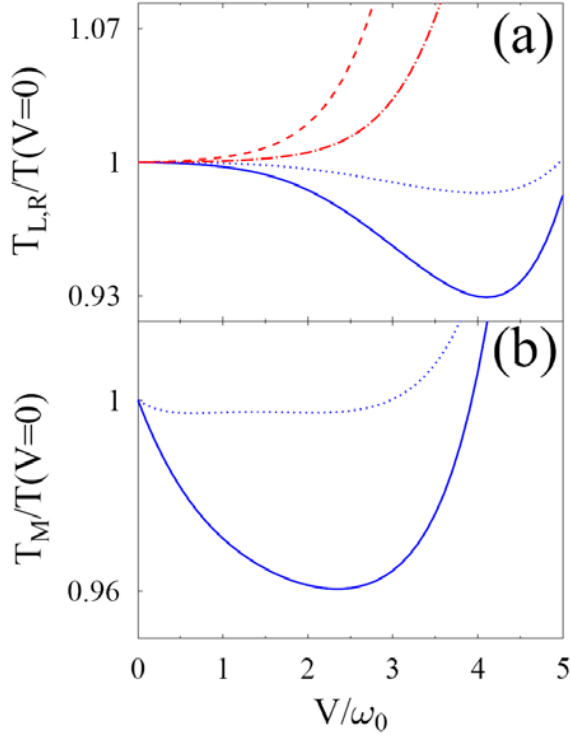

Figure 1: The effective non-equilibrium temperature $T_p$ vs. applied bias. Parameters of the calculation are $T = 300$ K, $d = 5$ A, $U = 0.2$ eV, and $\omega_p = 0.1$ eV. Shown are (a) Tempertures of the non-equilibrium parts of the electrodes in the M-I-M junction for the cases $T_{LL} = T_{RR} = T_{LR}^0$ (left electrode - solid line, blue; right electrode - dashed line, red) and $T_{LL} = T_{RR} = 10 \cdot T_{LR}^0$ (left electrode - dotted line, blue; right electrode – dash-dotted line, red) and (b) Temperature of the molecule asymmetrically coupled to the electrodes in a M-molecule-M junction. The molecular parameters of this calculation are $\varepsilon_0 = 0.3$ eV, $\Gamma_{ML} = 0.009$ eV, $\Gamma_{MR} = 0.001$ eV, $\omega_0 = 0.1$ eV (the same value of $\omega_0$ is used to normalize the voltage in Fig 1a), $M = 0.2$ eV, and $\Omega = 0$. The metallic rate constants $T_{LL} = T_{RR} = T_{LR}^0$ (full line; blue) and $T_{LL} = T_{RR} = 10 \cdot T_{LR}^0$ (dotted line; blue).

Eqs. (12) and (13) are used to determine the non-equilibrium distributions $f_L^{neq}(E)$ and $f_R^{neq}(E)$. The effective temperature $T_p$ can then be estimated by demanding that the heat flux to some probe vibration $\omega_p$ is zero

$$N_p \int_{-\infty}^{+\infty} f_K^{neq}(E - \omega_p)[1 - f_K^{neq}(E)] = (1 + N_p) \int_{-\infty}^{+\infty} f_K^{neq}(E + \omega_p)[1 - f_K^{neq}(E)] \quad (14)$$

where constant density of states in the contacts was assumed (wide-band approximation), and where

$$N_p = N_{BE}(\omega_p) = \left[\exp\left(\frac{\hbar\omega_p}{k_B T_p}\right) - 1\right]^{-1} \qquad (15)$$

is the Bose-Einstein distribution. It should be noted that the temperature estimate $T_p$ introduced by (14) and (15) depends on the choice of $\omega_p$. This is a consequence of the fact that, strictly speaking, temperature can not be introduced in a nonequilibrium system. Note also, that a more advanced (physically motivated) procedure for introducing an effective temperature can be used,[46]) however, for our purpose of qualitative illustration Eq.(14) will suffice.

Figure 1(a) shows the effective temperatures $T_p$ of the left and right nonequilibrium parts of the contacts vs. applied bias. Here positive bias means that the right electrode is biased positively relative to the left electrode, inducing electron current from left to right. While, as is expected, the right side heats up at positive bias, there is a range of bias voltages when left side of the junction cools down. It should be noted however that the effect is very sensitive to the relative timescales associated with $T_{LL}$, $T_{RR}$ as compared to $T_{LR}$ and is small unless the two electrodes are close enough to each other to make $T_{LR}$ large enough.

## 4. Metal-molecule-metal junction

Consider now a molecule (i.e. a species with internal degrees of freedom) lying in the junction and exchanging electrons and energy with the $L$ and $R$ electrodes. (between $L$ and $R$ electrodes). The molecule is coupled to the two non-equilibrium electronic reservoirs and to an equilibrium phonon bath that is taken to be at the ambient temperature. Its presence is assumed to cause only a small perturbation on the direct electron transport between the two metals and its effect on the non-equilibrium electron distributions at the two metal surfaces is disregarded. At steady-state the total energy flux to the molecule should be zero

$$J_{tot} = J_e + J_{ph} = 0 \qquad (16)$$

Here $J_e$ and $J_{ph}$ are electron and phonon mediated energy flux to the molecule[46]

$$J_e = \frac{1}{\hbar}\int_{-\infty}^{+\infty}\frac{dE}{2\pi}E\,\text{Tr}\Big\{\Big[\Sigma_{ML}^{<}(E)+\Sigma_{MR}^{<}(E)\Big]G^{>}(E) \\ -\Big[\Sigma_{ML}^{>}(E)+\Sigma_{MR}^{>}(E)\Big]G^{<}(E)\Big\} \quad (17)$$

$$J_{ph} = -\frac{1}{\hbar}\int_{0}^{\infty}\frac{d\omega}{2\pi}\omega\,\text{Tr}\Big\{\Pi^{<}(\omega)D^{>}(\omega)-\Pi^{>}(\omega)D^{<}(\omega)\Big\} \quad (18)$$

Here $G$ and $D$ are the Green functions (GFs) of the molecular electronic and vibrational degrees of freedom, $\Sigma_{MK}$ (K=L,R) is the molecular electronic self-energy (SE) due to coupling to the contact $K$, $\Pi$ is the molecular vibration SE due to coupling to an external thermal bath, and Tr{…} means trace over the molecular subspace. As usual $>$, $<$, $r$ and $a$ denote greater, lesser, retarded and advanced GFs and SEs. In what follows we consider a molecule characterized by one electronic level $m$ and one molecular vibration of freqecy $\omega_0$, (Eq. (5)), so the trace operation can be omitted.

Using the quasi-particle approximation for the molecular vibration and the Born approximation for the electron-vibration coupling, and describing the junction within the non-crossing approximation, one gets from (16) the steady state occupation $N_0$ of the molecular vibration $\omega_0$ (for detailed derivation see Refs. [46, 47])

$$N_0 = \frac{\Omega N_{BE}(\omega_0)+I_{+}}{\Omega+I_{+}-I_{-}} \quad (19)$$

where $N_{BE}$ is the Bose-Einstein equilibrium population at the ambient temperature and $\Omega = 2\pi\sum_{\beta}|U_{\beta}|^2\delta(\omega_0-\omega_{\beta})$ (assumed to be constant in the spirit of the wide-band approximation) is the damping rate of the molecular vibration due to coupling to the thermal bath, and where $I_{\pm}$ are given by

$$I_{\pm} \equiv M^2 \int_{-\infty}^{+\infty}\frac{dE}{2\pi}G_0^{<}(E)G_0^{>}(E\pm\omega_0) \quad (20)$$

in terms of the zero-order lesser and greater electronic GFs

$$G_0^{<}(E) = i\frac{\Gamma_{ML}(E)f_L^{neq}(E)+\Gamma_{MR}(E)f_R^{neq}(E)}{(E-\varepsilon_0)^2+(\Gamma_M/2)^2} \quad (21)$$

$$G_0^{>}(E) = -i\frac{\Gamma_{ML}(E)[1-f_L^{neq}(E)]+\Gamma_{MR}(E)[1-f_R^{neq}(E)]}{(E-\varepsilon_0)^2+(\Gamma_M/2)^2} \quad (22)$$

In Eqs. (21), (22) $\Gamma_{MK}(E) = 2\pi \sum_{k \in K_{neq}} |V_{mk}|^2 \delta(E - \varepsilon_k)$ are widths of the molecular level due to coupling to contacts $K = L, R$, and this coupling is assumed to be dominated by degrees of freedom belonging to the non-equilibrium parts of the corresponding metals. The effective temperature of the molecular vibration is then defined using the equilibrium form

$$T_M = \frac{\omega_0}{\ln[1 + 1/N_0]} \tag{23}$$

Clearly, a molecule coupled strongly to the cool ($L$) side of the junction described in Section 2.1 may cool down also, as is demonstrated in Fig. 1b.

In the situation just discussed the metals exchange charge and energy directly, and the effect of molecular presence on their non-equilibrium state is assumed insignificant. A more interesting situation is a metal-molecule-metal junction ($L_{eq} - L_{neq} - M - R_{neq} - R_{eq}$), where the coupling between the left and right contacts occurs only through the molecule. In this case molecular cooling is not simply due to coupling to the colder contact (made colder by direct coupling between the contacts), and is closely related to the transport process. The nonequilibrium steady-state of the junction may be described by rate equations analogous to (12), where now each nonequilibrium (surface) part of the electrode is coupled to its bulk equilibrium part on one side and to the molecule on the other (see Appendix A)

$$\frac{dN_L^{neq}(E,t)}{dt} = \frac{T_{LL}}{\hbar}\left[f_L^{eq}(E) - f_L^{neq}(E)\right] + \frac{1}{\hbar}\text{Tr}\left\{\Sigma_{ML}^>(E)G^<(E) - \Sigma_{ML}^<(E)G^>(E)\right\} = 0$$

$$\frac{dN_R^{neq}(E,t)}{dt} = \frac{T_{RR}}{\hbar}\left[f_R^{eq}(E) - f_R^{neq}(E)\right] + \frac{1}{\hbar}\text{Tr}\left\{\Sigma_{MR}^>(E)G^<(E) - \Sigma_{MR}^<(E)G^>(E)\right\} = 0 \tag{24}$$

where the electronic GFs are

$$G^{>,<}(E) = |G^r(E)|^2 \left[\Sigma_{ML}^{>,<}(E) + \Sigma_{MR}^{>,<}(E) + \Sigma_{ph}^{>,<}(E)\right] \tag{25}$$

To simplify our consideration we disregard the effect of coupling to the molecular vibration when evaluating the nonequilibrium electronic distributions in the contacts, i.e. we drop the self-energy associated with the electron-phonon coupling, $\Sigma_{ph}^{>,<}(E)$, when using (25) in (24). This approximation is used only to obtain the nonequilibrium distributions in the contacts. Once these distribution are determined, the calculation of the transport through molecular junction proceeds along the standard lines,[46, 48] taking the electron-phonon interaction into account and using the nonequilibrium distributions instead of the thermal (Fermi) ones in the metal leads. Eqs. (24) reduce to (12) with

$$T_{LR}(E) = \frac{\Gamma_{ML}(E)\Gamma_{MR}(E)}{(E-\varepsilon_0)^2 + (\Gamma_M(E)/2)^2} \qquad (26)$$

where $\Gamma_M(E) = \Gamma_{ML}(E) + \Gamma_{MR}(E)$. The energy dependence of $T_{LR}(E)$ now results from that of $\Gamma_{MK}(E)$. For the latter we again assume the square barrier-like behavior (13), i.e.,

$$\Gamma_{MK}(E) = \Gamma^0_{MK} \exp\left[-d\sqrt{(2m/\hbar^2)(U-E)}\right].$$

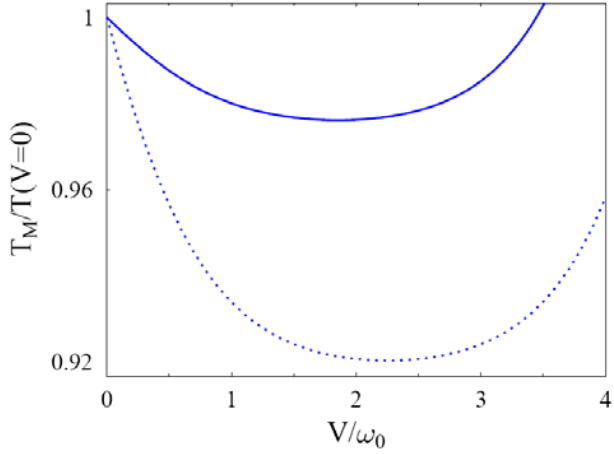

Figure 2: Effective molecular temperature, Eq.(14), vs. applied bias for metal-molecule-metal junction. The junction parameters were taken as in Fig. 1: $\varepsilon_0 = 0.3$ eV, $\omega_0 = 0.1$ eV, $M = 0.2$ eV, $d = 5$ A, $U = 0.2$ eV, $\omega_p = 0.1$ eV, and $\Omega = 0$. The metal-molecule coupling parameters were chosen $\Gamma^0_{ML} = 0.009$ eV and $\Gamma^0_{MR} = 0.001$ eV, and the metallic rate constants were taken $T_{LL} = T_{RR} = 4 \cdot 10^{-4}$ (full line; blue) and $T_{LL} = T_{RR} = 4 \cdot 10^{-3}$ (dotted line; blue).

Once the functions $f_K^{neq}$ ($K = L, R$) are evaluated from (24), the calculation proceeds as before and the effective molecular vibrational temperature is found from Eq. (14). Figure 2 demonstrates the possibility of molecular cooling also in this case. In this calculation, as well as in those described in the following section, the potential bias is taken to distribute symmetrically on the two molecule-metal contacts.

## 5. Inelastic cooling by coherent sub-resonance transport

It is well known that atomic cooling may be achieved by subresonance optical excitation.[49] Light of frequency slightly lower than the electronic transition energy may excite an atom by absorbing the needed extra energy from the kinetic energy "bath". A similar process can be used to cool molecular vibrations if electron-phonon[50] coupling is strong enough. In biased molecular junctions excitation by light is replaced by the electron flux, and the same mechanism that leads to inelastic tunneling may cause current induced vibrational cooling, provided that conditions that favor phonon absorption over phonon emission can be found. Such conditions turn out to be difficult to satisfy in most metal-molecule-metal junctions because the equivalent of the energy resolution provided by the optical excitation now arises from the voltage bias between to Fermi distributions and is limited by the temprature dependence of the latter. Indeed, we have found that the simplest molecular junction model comprising one available molecular electronic level coupled to one molecular oscillator and connected to two free-electron metals characterized by uniform state densities cannot be cooled in this way. Rather, to achieve cooling, a suitable structured spectral densities of the electronic baths is needed, as was recently emphasized in a slightly different context.[51] Very recently, an example of such system (an adatom bonded to an atomic wire, where the needed energy structure is provided by the mid-band dip in the transmission function of this system) was described by McEniry et al.[52] In what follows we describe two additional such possibilities.

**5.1 Inelastic cooling using semiconductor contacts**

In the system described below the required energy resolution is achieved by employing the sharp band-edge structure of semiconductor leads. Molecular junctions based on semiconductor substrates have recently attracted attention.[53-57] It was recently pointed out that negative differential resistance (NDR) may be observed in Si based molecular junctions due to a molecular electronic level passing by the Si band edge.[57] The presence of a gap between the valence and conduction band can in principle result in molecular cooling during conduction. For example, consider an intrinsic or p-doped semiconductor (below we use band structure typical to Si) biased such that the top of the valence band of the left electrode is below the bottom of the conduction band of the right electrode, with distance between their edges of order of $\omega_0$ (see scheme in the inset of Fig. 3a). In this case electron transition from left to right may be facilitated by absorption of energy from the molecular vibration, resulting in cooling of the vibration at some range of the bias voltage. Note that the oposite effect, electron tunneling with energy transfer to the vibration, is blocked here by the seminconductor band-gap.

The calculation reported below was performed within the NEGF approach outlined in Sect. 4 and described in Ref. [46]. It is now applied to the model (1)-(2), (4)-(7) without invoking a non-equilibrium layer near the electrodes-metal contacts. Also, asymmetry in the molecule-semiconductors coupling is not needed here. Instead, the needed structure in the energy dependence of the electron flux is achieved from the biased electronic structure of the semiconductor electrodes that are characterized by gaps in their densities of states, as seen in the inset of Fig.3. As before, the calculation yields the population of the molecular vibration coupled to the electronic system and the effective molecular vibrational temperature is found from (15).

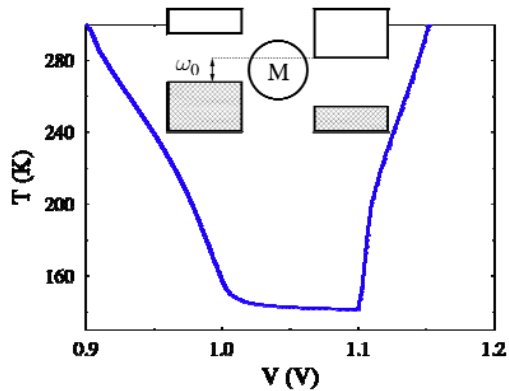

Figure 3: Effective temperature (determined from Eq. (23)) of a molecular vibration, displayed vs. applied bias of a molecular vibration in a silicon-molecule-silicon junction.

Figure 3 demonstrates this possibility in a semiconductor-molecule-semiconductor junction model, using a semiconductor bandgap of 1eV, typical to Si. The molecular level $\varepsilon_0$ is taken 0.4 eV above the valence band edge of the semiconductor in the unbiased junction. Other parameters used in this calculation are $\Gamma_{ML} = \Gamma_{MR} = 0.01\,\text{eV}$, $\omega_0 = 0.1\,\text{eV}$, $M = 0.2\,\text{eV}$, $\Omega = 5 \cdot 10^{-4}\,\text{eV}$ and $T$ (ambient temperature) = 300K. Again, the voltage bias is taken to distribute symmetrically on the two molecule-semiconductor bonds. Note that the top valence band edge of the left electrode alligns with the bottom conduction band edge of the right electrode at $V = 1\text{V}$, however cooling continues to be effective up to $V \sim 1.1\text{V}$ because of the energy structure provided by the molecular resonance $\varepsilon_0$. Above $V = 1.1\text{V}$ heating of the oscillator of $\omega_0 = 0.1\text{eV}$ becomes possible and finally the dominating process.

## 5.2 Sub-resonance transport in specific molecular structures

In the model described in the previous section the SC density of states, dominated by the pronounced (relative to $k_B T$) gap between the valence and conduction bands, provides the energy-structured spectral distribution that promotes phonon absorption in some voltage bias range. A suitable density of states can be also achieved in special cases of molecular electronic structure.

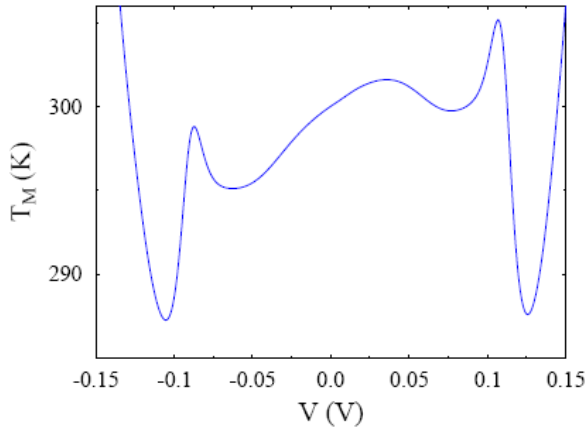

Figure 4: Effective molecular temperature vs. applied bias for 3-site bridge model. See text for parameters.

Figure 4 illustrates such possibility. The calculation follows again the procedure of Ref. [46] as outlined in Section 4. The molecule is represented by a 3-sites bridge, with on-site energies $-0.06$, $0.1$, and $-0.05$ eV (the Fermi energy $E_F$ of the unbiased junction is taken $0$) and nearest neighbor inter-site coupling $0.01$ eV. The leftmost and rightmost bridge levels are coupled to the left and right electrodes, respectively, with $\Gamma_{ML} = \Gamma_{MR} = 0.01$ eV. A single molecular vibration with $\omega_0 = 0.1$ eV is considered, with the electron-vibration coupling taken in the form

$$\sum_{m=0}^{3} M_{m,m+1} \hat{Q}_a (\hat{d}_m^\dagger \hat{d}_{m+1} + H.c.) \tag{27}$$

where $m = 0$ and $m = 4$ stand for the left and right electrodes, respectively and all coupling strengths $M$ are taken 0.2 eV. The electrodes equilibrium temperature of the contacts is $T = 300$ K. As before, the potential bias is assumed to fall symmetrically on the two electrodes. Specifically, in the calculation shown in Fig. 4 the left ($\varepsilon_1$) and right ($\varepsilon_3$) molecular levels of the bridge are assumed to shift with bias $V$ as $\varepsilon_1 + eV/2$ and $\varepsilon_3 - eV/2$, respectively. It should be emphasized that different ways by which V is distributed along the junction may lead to different temperature/bias behaviors.

## 6. Driven activation-induced cooling

The coherent cooling models described in Section 5 were seen to depend in a critical way on judicious choice of spectral densities, level structures and coupling strengths. Because some of these properties cannot be easily controlled, this makes realization of such cooling phenomena subject to accidental occurance of special circumstances. The activation-induced cooling process discussed below, essentially a classical analog of the mechanism discussed in Sect. 5 is of more general nature. Here electronic conduction is facilitated by thermal activation along the transport path. Specifically, if the electron experiences an uphill transition at some position along the transport path and if electron-phonon coupling is strong at that position the uphill current can be assisted by energy absorption from the phonon environment, cooling the latter in the process.

To demonstrate this process we consider a system described by a particular realization of the model (1)-(3). The Hamiltonian is written in the local electronic state represantation and comprises a nearest neighbor tight-binding electronic system with local coupling to a set of harmonic oscillators, with each oscillator coupled to one local electronic site[58] according to

$$\hat{H}_M = \sum_{j=1}^{N} \varepsilon_j \hat{d}_j^\dagger \hat{d}_j + u \sum_{j=1}^{N-1} \left( \hat{d}_{j+1}^\dagger \hat{d}_j + \hat{d}_j^\dagger \hat{d}_{j+1} \right) + \hbar\omega \sum_{j=1}^{N} \hat{a}_j^\dagger \hat{a}_j + \sum_{j=1}^{N} \lambda_j (\hat{a}_j^\dagger + \hat{a}_j) \hat{d}_j^\dagger \hat{d}_j \quad (28)$$

This system is coupled to several baths: each oscillator is taken to couple to its own thermal bath and the end electronic sites $j = 1, N$ are coupled to their respective free electron reservoirs, denoted $L$ and $R$, that represent metalic leads. The latter are taken to be at the same ambient temperature $T$, however the temperature $T_j$ of the thermal bath associated with system phonon $j$ is used to monitor local temperature change in the system as explained below. The electronic reservoirs are also characterized by their electronic chemical potentials, $\mu_L$ and $\mu_R$, whose difference represent potential bias. Note that the site energies $\varepsilon_j$ are not all equal. In particluar we will investigate the consequence of an uphill transition ($\varepsilon_{n+1} > \varepsilon_n$ for a left-to-right electronic current) on the non-equilibrium properties of this system and in particular on the local effective temperature $T_j$ assumed by each oscillator when a steady electronic current is driven trough the system by an imposed potential bias.

We will study this problem using a generalized master equation approach in the Markovian limit that relies on an assumed weak coupling between the system and the reservoirs and on the wide band approximation. This imposes no practical restriction on the description of thermal coupling between the system oscillators and their respective heat baths, however the treatment of electronic transport with this approach is applicable only to situations dominated by succesive metal-molecule-metal transitions rather than cotunneling processes. We note that a recent study by Prachar and Novotny[59] indicates problems in the calculation of current noise using such approach, however the computed average current seems to be valid.

Under the approximations outlined above, the markovian master equation for the time evolution of the reduced system density matrix $\rho_M$ under the effect of coupling to the electronic and thermal baths is (see Appendix B)

$$\frac{d\hat{\rho}_M}{dt} = -i\left[\hat{H}_M, \hat{\rho}_M\right] - (\mathcal{R}_{el}^L \hat{\rho}_M + \mathcal{R}_{el}^R \hat{\rho}_M) - \sum_{j=1}^{N} \mathcal{R}_{ph}^j (T_j)\hat{\rho}_M, \qquad (29)$$

where the Liouville superoperators $\mathcal{R}_{el}^{L/R}$ and $\mathcal{R}_{ph}^j(T_j)$ account respectively for the electron transfer interaction with the $L/R$ leads, and for the interaction of each molecular phonon $j$ with its associated thermal bath of temperature $T_j$. The superoperators $\mathcal{R}_{el}^K$, $K = L, R$ are given by

$$\mathcal{R}_{el}^K \hat{\rho} = \left(\gamma_{el}\left[\hat{d}_{j_K}, \hat{R}_+^K \hat{\rho}_M - \hat{\rho}_M \hat{R}_-^K\right] + \text{h.c.}\right), \quad K = L, R, \quad j_L = 1, j_R = N \quad (30)$$

where the operators $\hat{R}_{+/-}^K$ are defined by Eqs. (67) and (68). The superators affecting thermal coupling are given by

$$\mathcal{R}^j_{ph}(T_j)\hat{\rho}_M = \gamma_{ph}\left(\left[\hat{X}_j, \hat{R}^j_{ph}\hat{\rho}_M\right] + \text{h.c}\right); \quad j = 1,...,N \tag{31}$$

where $\hat{X}_j$ are system phonon operators that interact with the j-th phonon environment, (see Eq. (54); in the calculations presented below we have used $\hat{X}_j = \hat{a}^\dagger_j + \hat{a}_j$) and where the operators $\hat{R}^j_{ph}$ are given by Eqs. (70), (71). The rate parameters $\gamma_{el}$ and $\gamma_{ph}$ for the molecule-leads electron transfer and for the molecular oscillators – thermal bats energy transfer are determined by the corresponding spetral densities by standard golden-rule expressions (see Appendix B).

The numerical evaluation of the time evolution (29) is done in a truncated basis of eigenstates of the Hamiltonian $\hat{H}_M$. These eigenstates are calculated by diagonalizing the Hamiltonian matrix written in the basis of the Hamiltonian $\hat{H}_{M0}$ that does not include the electron-phonon interaction (last term in Eq. (28)). The truncation is done by representing each oscillator by a finite number of (lowest enegy) states. The evaluation of matrix elements such as $\langle k|\hat{\mathbf{O}}|l\rangle$ between exact system eigenstates, where $\hat{\mathbf{O}} = \hat{d}, \hat{d}^\dagger, \hat{a}, \hat{a}^\dagger$ is done by transforming back and forth between the $\hat{H}_M$ and $\hat{H}_{M0}$ bases.

Eq. (29) can be used to evaluate the currents going through the system. The electronic currents are obtained by multiplying both sides of this equation by the operator for the electron number on the system $\hat{N}_M = \sum_j \hat{d}^\dagger_j \hat{d}_j$:

$$\frac{\partial \text{Tr}(\hat{N}_M \rho_M)}{\partial t} = -\gamma_{el}\text{Tr}(\hat{N}_M \mathcal{R}^L_{el} \rho_M) - \gamma_{el}\text{Tr}(\hat{N}_M \mathcal{R}^R_{el} \hat{\rho}_M) \tag{32}$$

where $\text{Tr}(\hat{N}_M \mathcal{R}^{(ph)}_j(T_j)\hat{\rho}_M) = 0$ was used since the thermal reservoir cannot exchange electrons. Eq. (32) is the continuity equation with respect to charge number on the system and implies that the currents at the system interface with the metalic leads are given by

$$J_K = -\gamma_{el}\text{Tr}(\hat{N}_M \mathcal{R}_K \hat{\rho}_M); \quad K = L, R \tag{33}$$

These currents are equal at steady state. Similarly a continuity equation for the energy is obtained from

$$\frac{\partial \text{Tr}(\hat{H}_M \hat{\rho}_M)}{\partial t} = -\gamma_{el}\text{Tr}(\hat{H}_M \mathcal{R}^L_{el} \hat{\rho}_M) - \gamma_{el}\text{Tr}(H_M \mathcal{R}^R_{el} \hat{\rho}_M) - \gamma_{ph}\sum_j \text{Tr}(\hat{H}_M \mathcal{R}^j_{ph}(T_j)\hat{\rho}_M). \tag{34}$$

From this, the heat flux between the $j$-th system phonon and its thermal reservoir is naturally identified as

$$\dot{Q}_j = -\gamma_{ph} \text{Tr}(\hat{H}_M \mathcal{R}_{ph}^j(T_j)\hat{\rho}_M). \tag{35}$$

A positive $\dot{Q}_j$ means that heat flows into the system through the oscillator $j$.

In general, the steady state temperature of the system phonons reflects the balance of energy exchange with the electronic subsystem and the external thermal bath(s). In the present calculations we choose $\gamma_{ph}$ small enough so it does not affect the steady-state temperature of the system phonons. Instead, coupling to the thermal reservoirs is used to determine the local temperature of any system oscillator $j$ by a method introduced (using the NEGF formalism) in Ref. [46] – as the temperature $T_j$ of the corresponding thermal reservoir for which the energy current $\dot{Q}_j$ vanishes. To this end we supplement Eq. (28) by artificial dynamical equations for the baths temperatures $T_j$ that drives these temperatures to their zero currents values

$$\frac{\partial T_j}{\partial t} = -\nu \dot{Q}_j \quad (\nu > 0) \tag{36}$$

At steady state $\dot{Q}_j = 0$, and $T_j$ is identified as the effective temperature of the oscillator $j$. The choice of the relaxation parameter $\nu$ is arbitrary and subjected only to numerical convergence considerations. Fig. 5 shows a typical time evolution obtained from Eqs. (28) and (36). Here we use a model with two electronic sites, $N = 2$ (implying four electronic states) and employ the 4 lowest levels of each oscillator. Our Hilbert space is therefore of dimension $4 \times 4^2 = 64$. This calculation was done with the parameters (in units of $\hbar\omega$) $\varepsilon_1 = 0$, $\varepsilon_2 = 1$ (measured from the leads Fermi energy in the unbiased junction), $u = 1$, $\lambda_1 = \lambda_2 = 1$, $\gamma_{el} = 0.1$, $\gamma_{ph} = 0.01$ and $k_B T = 1$. Note that for our purpose the choice of $\gamma_{ph}$ is immaterial, except that it should be chosen small enough not to affect the final tempertuare $T_j, j = 1,2$.[60] In addition, $w$ (defined by Eq. (64)) is taken large enough so as not to affect the calculated results and to make them insensitive to the choice of $\varepsilon_0$ (Eq. (64)) when the latter is taken in the vicinity of $\varepsilon_1$ and $\varepsilon_2$. In the calculations presented below we have used $w = 80$ eV and $\varepsilon_0 = 0$. Ohmic spectral densities are taken to characterise the phonon baths. Fig. 5 shows the time-evolution of the local temperatures $T_1$ and $T_2$ and (in the inset) of the heat currents, $\dot{Q}_1, \dot{Q}_2$ between the system and the two heat reservoirs. As exlained above, this time evolution, which does depend on the choice of $\gamma_{ph}$, does not reflect a physical process, only the progress of the computation. In the long time limit $\dot{Q}_j$ vanishes and the local temperatures $T_j$ reach their steady-state values.

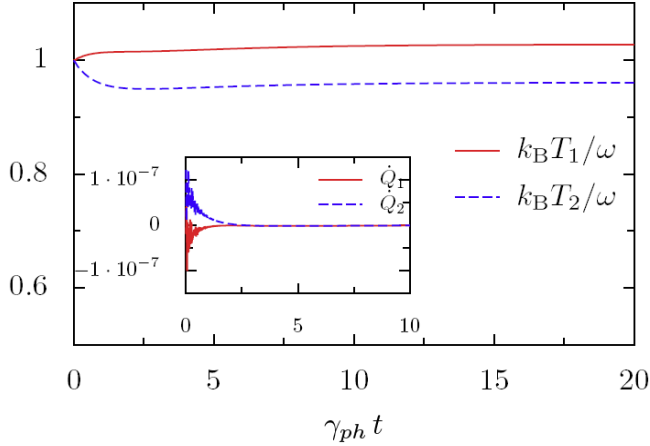

Figure 5: Time-evolution of local temperatures $T_1$ and $T_2$ for $(\mu_L, \mu_R) = (1.0, 0)$. The inset shows the time evolution of the heat currents $\dot{Q}_j$ that vanish in the long time limit. This automatically determines the local temperature $T_j$ at the steady state. The temperature on both leads are $k_B T / \omega = 1$.

The three panels of Figure 6 show the steady state local temperatures $T_1, T_2$ obtained for the same two-state model and plotted against the bias voltage $\mu_L - \mu_R$. In this calculation we have used $\varepsilon_1$=0.8 eV, $\varepsilon_2$=1.0 eV, $\varepsilon_0$ = 0, $k_B T$ = 0.3 eV and $\mu_R$=0, while $\omega$ and $\mu_L$ vary as shown. Other parameters are as in Fig. 5. We see that the local temperature on site 2 decreases in the positive bias regime and that the effect is stronger for lower oscillator frequency.

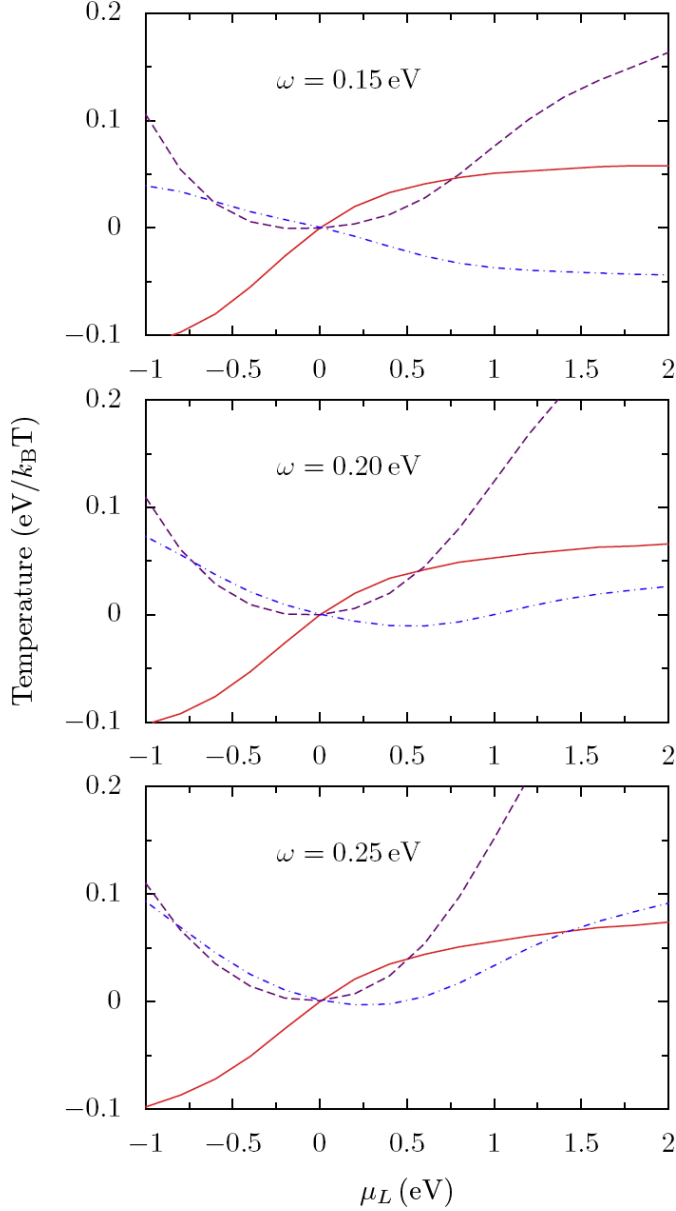

Figure 6: Local temperatures, $T_j - T$; $j = 1,2$ displayed against the bias voltage $\mu_L - \mu_R$ ($\mu_R = 0$). Also shown is electronic current $J_c$. Cooling is indicated by negative values of $T_j - T$. Dashed line (purple): $T_1 - T$. Dashed-dotted line (blue): $T_2 - T$. Solid line (red) – $J_c$ (x100 in units of $e^2V$ ($\hbar = 1$))

## 7. Summary and conclusion

Understanding cooling phenomena in non-equilibrium nanojunctions, in particular molecular junctions, is important in considerations of their stability and functionality. In this paper we have studied, within simple models, different mechanisms for junction cooling under applied bias. One such mechanism (Sections 3 and 4) is associated with cooling of the high voltage side of a 2-terminal junction due to depletion of high-energy electrons, which establishes a non-equilibrium electronic distribution that corresponds to lower effective temperature. A molecule that couples more strongly to this side of the junction than to the other, hotter, side will be cooled under the applied bias. This can happen irrespective of whether the molecule actively carries the junction current or participate in the electrical conduction process merely as a spectator.

A second mechanism (Section 5) considered is the inelastic electron tunneling equivalent of laser cooling. When the available electronic energy falls short of what is needed for efficient transmission, the needed excess energy can be provided by absorbing phonons from the vibrational subsystem, cooling the the later. Actual cooling requires the suppression of the usually more efficient phonon emission processes, which can be achieved for particular energy structures of the electron transmission. Such structure can be realized in molecular-semiconductor junctions, when the phonon emission channel is closed by the semiconductor band gap, and also in particular resonant structures of the molecular bridge. Other possible realizations of this concept have been recently discussed by other groups[51, 52] and an interesting application was demonstrated.[51]

When a current is driven through a system where current flow requires a local uphill motion, energy absorption from the local environment make take place, resulting in cooling (Section 6). This is in principle similar to the mechanism just discussed, but under strong coupling with the environment it becomes less restrictive and independent of the particlular energy dependence of the transmission process. In fact, a classical limit of this process can be envisualized.[61]

It should be pointed out the the models discussed in this paper where analyzed in the absence of electron-electron interactions. Such interactions will affects in particular systems sustaining large electrical currents where issues of cooling (or reduction of heating) are of particular interest. Understanding such effects, as well as extending our calculations to models of realistic systems will be the next steps of our study.

## Appendix A

Here we derive Eqs (12) and (24). Let consider continuum of states $\{k\}$ coupled to a system $\{m\}$ whose nature is defined below. The continuum is assumed to be a set of free electronic states. Following procedure used to derive general expression for current within NEGF[62], one gets for rate of population change in the state $k$

$$\frac{dn_k(t)}{dt} \equiv <\hat{c}_k^\dagger(t)\hat{c}_k(t)> = -\frac{i}{\hbar}\sum_{m,m'}V_{mk}V_{km'}\int_{-\infty}^{t}dt_1$$

$$\left\{(1-n_k(t))\left[G_{m'm}^<(t,t_1)e^{i\varepsilon_k(t-t_1)/\hbar} + G_{m'm}^<(t_1,t)e^{i\varepsilon_k(t_1-t)/\hbar}\right]\right.$$

$$\left. +n_k(t)\left[G_{m'm}^>(t,t_1)e^{i\varepsilon_k(t-t_1)/\hbar} + G_{m'm}^>(t_1,t)e^{i\varepsilon_k(t_1-t)/\hbar}\right]\right\} \quad (37)$$

where the following equations for free lesser and greater GFs of the continuum where used

$$g_k^<(t) = \frac{i}{\hbar}n_k(t)e^{-i\varepsilon_k t/\hbar}; \quad g_k^>(t) = -\frac{i}{\hbar}[1-n_k(t)]e^{-i\varepsilon_k t/\hbar} \quad (38)$$

Summing (37) over states $k$ with energy $\varepsilon_k = E$ and introducing density of states $A_K(E)$ and occupation at energy $E$ $f_K(t,E)$

$$N_K(t,E) \equiv f_K(t,E)A_K(E) = 2\pi\sum_{k\in K}n_k(t)\delta(E-\varepsilon_k) \quad (39)$$

leads to

$$\frac{dN_K(t,E)}{dt} = -\frac{i}{\hbar}\sum_{m,m'}\int_{-\infty}^{t}dt_1\,\Gamma_{mm'}^K(E)$$

$$\left\{[1-f_K(t,E)][G_{m'm}^<(t,t_1)e^{iE(t-t_1)/\hbar} + G_{m'm}^<(t_1,t)e^{iE(t_1-t)/\hbar}]\right. \quad (40)$$

$$\left. +f_k(t,E)[G_{m'm}^>(t,t_1)e^{iE(t-t_1)/\hbar} + G_{m'm}^>(t_1,1)e^{iE(t_1-t)/\hbar}]\right\}$$

where

$$\Gamma_{mm'}^K(E) \equiv 2\pi\sum_k V_{mk}V_{km'}\delta(E-\varepsilon_k) \quad (41)$$

Now we consider two choices of the system. If the system is a set of free electronic states (continuum of states) then

$$G_{mm'}^{>,<}(t_1,t_2) = \delta_{m,m'}g_m^{>,<}(t_1-t_2) \quad (42)$$

and $g_m^{>,<}$ satisfies (38). Eq. (40) reduces to

$$\frac{dN_K(t,E)}{dt} = \frac{1}{\hbar} T_{KM}(E) \left[ f_M(t,E) - f_K(t,E) \right] \tag{43}$$

where

$$T_{KM}(E) \equiv \sum_{m \in M} \Gamma^K_{mm}(E) \delta(E - \varepsilon_m) = 2\pi \sum_{k \in K, m \in M} |V_{km}|^2 \delta(E - \varepsilon_k) \delta(E - \varepsilon_m) \tag{44}$$

Under wide-band approximation one gets (12) for the case when the continuum $K$ is coupled to two different continua.

When the system is a molecule, the single electron states used as a basis are not eigenstates of the molecular Hamiltonian. Eq. (40) can then be simplified only in the steady-state situation, when $G^{>,<}_{mm'}(t_1, t_2) = G^{>,<}_{mm'}(t_1 - t_2)$. In this case one gets

$$\frac{dN_K(t,E)}{dt} = 0 = \frac{1}{\hbar} Tr \left[ \Sigma^>(E) G^<(E) - \Sigma^<(E) G^>(E) \right] \tag{45}$$

where

$$\Sigma^<_{mm'}(E) = i f_K(E) \Gamma^K_{mm'}(E); \quad \Sigma^>_{mm'}(E) = -i[1 - f_K(E)] \Gamma^K_{mm'}(E) \tag{46}$$

and $\Gamma^K_{mm'}$ is defined in (41). Consideration of a continuous manifold of states coupled to another continuum on one side and to the molecule on the other leads to (24).

## Appendix B

For a system (M) interacting with one or more heat baths (K), let the Hamiltonian be

$$\hat{H} = \hat{H}_M + \sum_K \hat{H}_K + \sum_K \hat{H}_{MK} \tag{47}$$

with

$$\hat{H}_{MK} = \sum_n \hat{X}^{Kn} \hat{Y}^{Kn} \tag{48}$$

In Eq. (48) $\hat{X}$ and $\hat{Y}$ are system and bath operators, respectively. Each bath is taken to be in its own equilibrium but different baths can be characterized by different equilibrium properties (e.g. temperatures and/or chemical potentials). The time evolution of the system density operator $\hat{\rho} = \text{Tr}_{\text{all baths}}(\hat{\rho}^{tot})$ in the Redfield (weak system bath coupling and the Markovian limit) approximation is given, in the representation of eigenstates of $\hat{H}_M$ by (see, e.g. [63, 64])

$$\frac{d\rho_{ab}(t)}{dt} = -i\omega_{ab}\rho_{ab} - \sum_{cd}\Big(R_{ac,cd}(\omega_{dc})\rho_{db}(t) + R^*_{bd,dc}(\omega_{cd})\rho_{ac}(t) \tag{49}$$
$$-\big[R_{db,ac}(\omega_{ca}) + R^*_{ca,bd}(\omega_{db})\big]\rho_{cd}(t)\Big)$$

where the indices $i = a,...,d$ denotes eigenstates of $\hat{H}_M$, $\hat{H}_M|i\rangle = E_a|i\rangle$, and $\omega_{i,i'} = (E_i - E_{i'})/\hbar$,[1]

$$R_{ab,cd}(\omega) \equiv \int_0^\infty dt\, M_{ab,cd}(t) e^{i\omega t} \tag{50}$$

$$M_{ab,cd}(t) = \frac{1}{\hbar^2}\sum_K \sum_{m,n} C^K_{mn}(t) X^{Km}_{ab} X^{Kn}_{cd} \tag{51}$$

In Eq. (51) $X^{Km}_{ab} = \langle a|\hat{X}^{Km}|b\rangle$, and $C^K_{mn}(t)$ are bath correlation functions

$$C^K_{mn}(t) = \langle \hat{Y}^{Km}(t)\hat{Y}^{Kn}(0)\rangle \tag{52}$$

In our model, the system (Eq. (28)) comprises $N$ electronic levels and $N$ harmonic oscillators. Electronic levels 1 and $N$ are coupled to free electron reservoirs $L$ and $R$, respectively, by the electron transfer interaction

$$\hat{H}_{ML} = \hat{d}_1^\dagger \sum_{k \in L} V^L_{1k} \hat{c}_k + \text{h.c}$$
$$\hat{H}_{MR} = \hat{d}_N^\dagger \sum_{k \in R} V^R_{Nk} \hat{c}_k + \text{h.c} \tag{53}$$

while each system oscillator $j$ is coupled to its own heat bath

$$\hat{H}_{Mj} = \hat{X}_j \hat{Y}_j \tag{54}$$

where $\hat{X}_j$ in an operator in the system oscillator space and $\hat{Y}_j$ is an operator in the corresponding thermal bath of temperature $T_j$. In our application these thermal baths are taken as identical free boson reservoirs with $\hat{Y}_j$ linear in the corresponding boson coordinates, characterized by an ohmic spectral density.

---

[1] More generally, these are eigenstates of $\hat{H}_M + \sum_K \langle \hat{H}_{MK}\rangle$ where $\langle \hat{H}_{MK}\rangle$ are averages over the respective baths. In the models used in this paper the latter are zero.

The overall system-baths interaction is thus $\sum_K \hat{H}_{MK}$ with $K = L, R, \{j\}$. The structure of Eqs. (49)-(52) (which reflects the level of approximation at which the master equation is obtained) implies that these baths contribute additively to the master equation

$$\frac{\partial \hat{\rho}(t)}{\partial t} + \frac{i}{\hbar}\left[\hat{H}_M, \hat{\rho}(t)\right] = \mathcal{R}_{el}\hat{\rho} + \mathcal{R}_{ph}\hat{\rho}. \tag{55}$$

$$\mathcal{R}_{el} = \mathcal{R}_{el}^L + \mathcal{R}_{el}^R; \quad \mathcal{R}_{ph} = \sum_j \mathcal{R}_{ph}^j \tag{56}$$

Consider for example $\mathcal{R}_{el}^L$. For the coupling $\hat{H}_{ML}$ between the system and the left electronic reservoir we have in Eq. (48) $\hat{X}^{L1} = \hat{d}_1$; $\hat{Y}^{L1} = \sum_{k \in L} V_{k1}^L \hat{c}_k^\dagger$; $\hat{X}^{L2} = \hat{d}_1^\dagger$; $\hat{Y}^{L2} = \sum_{k \in L} V_{1k}^L \hat{c}_k$.[2] The relevant correlation functions are

$$C_{12}^L(t) = \sum_{k \in L} |V_{1k}^L|^2 n_L(\varepsilon_k) e^{i(\varepsilon_k/\hbar)t} = \int_{-\infty}^{\infty} d\varepsilon\, g_1(\varepsilon) n_L(\varepsilon) e^{i(\varepsilon/\hbar)t} \tag{57}$$

$$C_{21}^L(t) = \sum_{k \in B} |V_{1k}^L|^2 (1 - n_L(\varepsilon_k)) e^{-i(\varepsilon_k/\hbar)t} = \int_{-\infty}^{\infty} d\varepsilon\, g_1^L(\varepsilon)(1 - n_L(\varepsilon)) e^{-i(\varepsilon/\hbar)t} \tag{58}$$

where ($\mu_L$ is the chemical potential in $L$)

$$n_L(\varepsilon) = \frac{1}{e^{\beta(\varepsilon - \mu_L)} + 1}; \quad 1 - n_L(\varepsilon) = n_L(-\varepsilon) \tag{59}$$

and where $g_1^L(\varepsilon)$ is the coupling weighted density of states for electronic level 1 interacting with the left electronic bath $L$, defined by

$$\sum_k |V_{1k}^L|^2 f(\varepsilon_k) = \int_{-\infty}^{\infty} d\varepsilon\, f(\varepsilon) \sum_k |V_{1k}^L|^2 \delta(\varepsilon - \varepsilon_k) = \int_{-\infty}^{\infty} d\varepsilon\, f(\varepsilon) g_1^L(\varepsilon) \tag{60}$$

for any function $f(\varepsilon)$. In Eqs. (57), (58) we have assumed that the edges of the metal band do not affect the dynamics and extended the limits of integration to infinity.

The contribution of $C_{12}^L(t)$ to the R elements in Eq. (49) is

---

[2] Some care should in principle be taken in consideration of the fact that, these being fermionic operators, $\hat{X}$ and $\hat{Y}$ do not commute. In the present formulation this has no consequence because needed permutations are even.

$$R_{ab,cd}(\omega) = \ldots + \left(\hat{d}_1^\dagger\right)_{ab} \left(\hat{d}_1\right)_{cd} \int_0^\infty dt\, C_{12}^L(t) e^{i\omega t} + \ldots \tag{61}$$

Consider the Fourier transform involved:

$$\int_0^\infty dt\, C_{12}^B(t) e^{i\omega t} = \int_0^\infty dt\, e^{i\omega t} \int_{-\infty}^\infty d\varepsilon\, g_1^L(\varepsilon) n_L(\varepsilon) e^{i(\varepsilon/\hbar)t} \tag{62}$$

In the absence of the Fermi function $n_L(\omega)$ or when the molecule-metal coupling is small relative to $\beta^{-1} = k_B T$, Eq. (62) can be approximated by $(1/2)\Gamma_1^L n_L(\omega)$, where $\Gamma_1^L = g_1^L(\omega)/2\pi$ is assumed energy independent (wide band approximation). More generally, we follow the procedure of refs [65, 66] and represent $g_j^K(\varepsilon)$ as a sum over overlapping Lorentzians

$$g_j^K(\varepsilon) = \frac{\Gamma_j^K}{2\pi} \sum_v \frac{w_v^2}{(\varepsilon - \varepsilon_v)^2 + w_v^2}; \quad (j, K) = (1, L) \text{ or } (N, R) \tag{63}$$

and evaluate the integral (62) by complex integration. In the calculations done in this paper we have taken all $\Gamma$ equal and have used a single Lorentzian

$$g(\varepsilon) = \frac{\gamma_{el}}{2\pi} \frac{w^2}{(\varepsilon - \varepsilon_0)^2 + w^2} \tag{64}$$

with width $w$ large enough (relative to all other energy scales; $d$ may be considered as representing the metal bandwidth) so that the choice of the center $\varepsilon_0$ has no consequence. Eq. (62) can then be evaluated using the identity

$$\frac{1}{2\pi} \int_0^\infty dt\, e^{i\omega t/\hbar} \int_{-\infty}^\infty d\varepsilon \frac{w^2}{(\varepsilon - \varepsilon_0)^2 + w^2} \frac{e^{i\varepsilon t/\hbar}}{e^{\beta(\varepsilon - \mu)} + 1} = \frac{\hbar w}{2} \frac{1}{[w - i\hbar(\omega + \varepsilon_0)][e^{\beta(iw - \mu + \varepsilon_0)} + 1]}$$
$$+ \hbar \beta^{-1} \sum_{j=0}^\infty \frac{w^2}{w^2 + [\mu - \varepsilon_0 + i\pi\beta^{-1}(2j+1)]^2} \frac{1}{\mu + \hbar\omega + i\pi\beta^{-1}(2j+1)} \tag{65}$$

where in the r.h.s. $w$ is positive.

Similar considerations apply to all contributions to $\mathcal{R}_{el}$, and finally lead to the following general form of the electronic Liouville operator

$$\frac{\partial \rho(t)}{\partial t} + \frac{i}{\hbar}[H_M, \rho(t)] = -\sum_{K=R,L} \Gamma^K \left\{ \left[d, R_+^K \rho(t)\right] + \left[d, R_+^K \rho(t)\right]^\dagger + \left[\rho(t) R_-^K, d\right] + \left[\rho(t) R_-^K, d\right]^\dagger \right\}$$

with (for the spectral function (64))

$$\langle k | R_+ | \ell \rangle = \langle k | d^\dagger | \ell \rangle \left[ \frac{\hbar w}{2} \frac{1}{[w+i(E_k - E_\ell - \varepsilon_0)][e^{\beta(iw - \mu + \varepsilon_0)} + 1]} \right.$$

$$\left. + \hbar \beta^{-1} \sum_{j=0}^{\infty} \frac{w^2}{w^2 + [\mu - \varepsilon_0 + i\pi\beta^{-1}(2j+1)]^2} \frac{1}{[\mu - (E_k - E_\ell) + i\pi\beta^{-1}(2j+1)]} \right]$$

(67)

$$\langle k | R_- | \ell \rangle = \langle k | d^\dagger | \ell \rangle \left[ \frac{\hbar w}{2} \frac{1}{[w+i(E_k - E_\ell - \varepsilon_0)][e^{-\beta(iv_D - \mu + \varepsilon_0)} + 1]} \right.$$

$$\left. - \hbar \beta^{-1} \sum_{j=0}^{\infty} \frac{w^2}{w^2 + [\mu - \varepsilon_0 + i\pi\beta^{-1}(2j+1)]^2} \frac{1}{[\mu - (E_k - E_\ell) + i\pi\beta^{-1}(2j+1)]} \right]$$

(68)

Next consider the phonon term $\mathcal{R}_{ph}^j \hat{\rho}$. For our purpose the simplest choice for system oscillator-thermal environment coupling is sufficient. The thermal bath $j$ is taken as a system of independent bosons, with corresponding raising and lowering operators $\hat{b}_l^{(j)\dagger}$ and $\hat{b}_l^{(j)}$ and the interaction Hamiltonian is taken as

$$\hat{H}_{Mj} = \lambda \hat{X}_j \sum_l \left( \beta_{jl} \hat{b}_l^{(j)} + \beta_{jl}^* \hat{b}_l^{(j)\dagger} \right) \tag{69}$$

which is obviously of the form (54). It is also sufficient to take the simplest choice $\hat{X}_j = \hat{a}_j^\dagger + \hat{a}_j$, which we do in the present calculation. The general results (49)-(52) then lead to

$$\mathcal{R}_{ph}^j (T_j) \hat{\rho}_M = \gamma_{ph} \left( \left[ \hat{X}_j, \hat{R}_{ph}^j \hat{\rho}_M \right] + \left[ \hat{X}_j, \hat{R}_{ph}^j \hat{\rho}_M \right]^\dagger \right) \tag{70}$$

$$\langle k | \hat{R}_{ph}^j | \ell \rangle = \text{sgn}(E_k - E_\ell) J_j(|E_k - E_\ell|) \langle k | \hat{X}_j | \ell \rangle n_j(E_k - E_\ell), \tag{71}$$

where

$$n_j(E) = \frac{1}{e^{E/k_B T_j} - 1} \tag{72}$$

where $\gamma_{ph} = |\lambda|^2$ and $J_j(E)$ is the spectral function for the coupling with bath $j$

$$J_j(E) = \frac{1}{\pi} \sum_l |\beta_{jl}|^2 \delta(E - \hbar\omega_{jl}) \tag{73}$$

(66)

with $\omega_{jl}$ being the frequency for the $l$ th mode at the jth bath.

## Acknowledgements

M.G. gratefully acknowledges support from the UCSD Startup Fund and the UC Academic Senate research grant. A.B. is supported by BES and LDRD funds at Los Alamos. KS was supported by MEXT (Grant no. 21740288). The research of AN is supported by the European Science Council (FP7 /ERC grant no. 226628), the German-Israel Foundation, the Israel – Niedersachsen Research Fund, the Lion Foundation and the Israel Science Foundation. This paper was concluded during AN's visit to the University of Konstanz, supported by the Humboldt Foundation. AN thanks W. Dieterich (University of Konstanz) for many helpful discussions during that time.

## References

[1] K. Schwab, E. A. Henriksen, J. M. Worlock, et al., Nature **404**, 974 (2000).
[2] P. Kim, L. Shi, A. Majumdar, et al., Phys. Rev. Letters **87**, 215502 (2001).
[3] D. Cahill, K. Goodson, and A. Majumdar, Journal of Heat Transfer **124**, 223 (2002).
[4] L. Shi and A. Majumdar, Journal of Heat Transfer **124**, 329 (2002).
[5] D. Cahill, W. K. Ford, K. E. Goodson, et al., Journal of Applied Physics **93**, 793 (2003).
[6] A. P. van Gelder, A. G. M. Jansen, and P. Wyder, Physical Review B **22**, 1515 (1980).
[7] R. K. Lake and S. Datta, Phys. Rev. B **45**, 6670 (1992).
[8] R. K. Lake and S. Datta, Phys. Rev. B **46**, 4757 (1992).
[9] B. N. J. Persson and P. Avouris, Surface Science **390**, 45 (1997).
[10] M. J. Montgomery, T. N. Todorov, and A. P. Sutton, J Phys.: Cond. Matter **14**, 5377 (2002).
[11] L. G. C. Rego and G. Kirczenow, Phys. Rev. Letters **81**, 232 (1998).
[12] K. R. Patton and M. R. Geller, Phys. Rev. B **64**, 155320 (2001).
[13] D. Segal, A. Nitzan, and P. Hanggi, J. Chem. Phys. **119**, 6840 (2003).
[14] D. Segal and A. Nitzan, Phys. Rev. Lettters **94**, 034301 (2005).
[15] D. Segal and A. Nitzan, J. Chem Phys. **122**, 194704 (2005).
[16] D. Segal and A. Nitzan, J. Chem. Phys. **117**, 3915 (2002).
[17] A. Cummings, M. Osman, D. Srivastava, et al., Physical Review B (Condensed Matter and Materials Physics) **70**, 115405 (2004).
[18] Y. C. Chen, M. Zwolak, and M. Di Ventra, Nano Lett. **3**, 1691 (2003).

[19] I. Paul and G. Kotliar, Physical Review B (Condensed Matter and Materials Physics) **67**, 115131 (2003).

[20] A. Y. Smirnov, L. G. Mourokh, and N. J. M. Horing, Phys. Rev. B **67**, 115312 (2003).

[21] N. M. Chtchelkatchev, W. Belzig, and C. Bruder, Phys. Rev. B **70**, 193305 (2004).

[22] J. Koch and F. v. Oppen, Phys. Rev. Letters **94**, 206804 (2005).

[23] J. Koch and F. von Oppen, Physical Review B (Condensed Matter and Materials Physics) **72**, 113308 (2005).

[24] J. Koch, M. Semmelhack, F. v. Oppen, et al., Phys. Rev. B **73**, 155306 (2006).

[25] T. Frederiksen, M. Brandbyge, N. Lorente, et al., Phys. Rev. Letters **93**, 256601 (2004).

[26] M. Paulsson, T. Frederiksen, and M. Brandbyge, Phys. Rev. B **72**, 201101 (2005).

[27] M. Paulsson, T. Frederiksen, and M. Brandbyge, Nano Letters **6**, 258 (2006).

[28] D. Segal, The Journal of Chemical Physics **128**, 224710 (2008).

[29] D. Segal, Physical Review Letters **101**, 260601 (2008).

[30] L.-A. Wu and D. Segal, Physical Review Letters **102**, 095503 (2009).

[31] W. C. Lo, L. Wang, and B. W. Li, Journal of the Physical Society of Japan **77** (2008).

[32] Z. Liu and B. Li, Physical Review E (Statistical, Nonlinear, and Soft Matter Physics) **76**, 051118 (2007).

[33] B. Li, L. Wang, and G. Casati, Physical Review Letters **93**, 184301 (2004).

[34] F. Pistolesi, Journal of Low Temperature Physics **154**, 199 (2009).

[35] J. P. Pekola, T. T. Heikkilä, A. M. Savin, et al., Physical Review Letters **92**, 056804 (2004).

[36] F. Giazotto, T. T. Heikkila, A. Luukanen, et al., Reviews of Modern Physics **78**, 217 (2006).

[37] J. P. Pekola and F. W. J. Hekking, Physical Review Letters **98**, 210604 (2007).

[38] S. Rajauria, P. S. Luo, T. Fournier, et al., Physical Review Letters **99**, 047004 (2007).

[39] O.-P. Saira, M. Meschke, F. Giazotto, et al., Physical Review Letters **99**, 027203 (2007).

[40] F. W. J. Hekking, A. O. Niskanen, and J. P. Pekola, Physical Review B (Condensed Matter and Materials Physics) **77**, 033401 (2008).

[41] Z. Ioffe, T. Shamai, A. Ophir, et al., Nature Nanotech. **3**, 728 (2008).

[42] S. Zippilli, G. Morigi, and A. Bachtold, Physical Review Letters **102**, 096804 (2009).

[43] L. D. Landau and E. M. Lifshits, *Quantum Mechanics: Non-relativistic Theory* (Pergamon Press, New York, 1965).

[44] J. G. Simmons, J. Appl. Phys. **34**, 1793 (1963).

[45] C. Aguilera-Navarro, H. Iwamoto, and V. M. d. Aquino, Int. J. Theor. Phys. **43**, 483 (2004).

[46] M. Galperin, M. Ratner, and A. Nitzan, Phys. Rev. B **75**, 155312 (2007).

[47] M. Galperin, M. A. Ratner, and A. Nitzan, The Journal of Chemical Physics **130**, 144109 (2009).

[48] M. Galperin, M. A. Ratner, and A. Nitzan, J. Chem. Phys. **121**, 11965 (2004).

[49] D. Leibfried, R. Blatt, C. Monroe, et al., Reviews of Modern Physics **75**, 281 (2003).
[50] In this paper we use "phonon" to describe all types of nuclear oscillations, including molecular vibrations.
[51] J. R. Prance, C. G. Smith, J. P. Griffiths, et al., Physical Review Letters **102**, 146602 (2009).
[52] E. J. McEniry, T. N. Todorov, and D. Dundas, Journal of Physics-Condensed Matter **21**, 5 (2009).
[53] T. Rakshit, G.-C. Liang, A. W. Ghosh, et al., Nano Letters **4**, 1803 (2004).
[54] R. A. Wolkow, Jpn. J. Appl. Phys. **40**, 4378 (2001).
[55] S. Lenfant, C. Krzeminski, C. Delerue, et al., Nano Lett. **3**, 741 (2003).
[56] M. Hersam and R. Reifenberger, MRS Bulletin **29**, 385 (2004).
[57] T. Rakshit, G. C. Liang, A. W. Ghosh, et al., Physical Review B **72**, 125305 (2005).
[58] Note that the electron-phonon coupling involving the oscillator j in Eq. (28) is manifested in any electron transfer event involving that site, and is therefore a property of site j as well as its nearest neighbors.
[59] J. Prachar and T. Novotn´y, cond-mat/0902.2382 (2009).
[60] In practical applications it is advantageous to choose $\gamma_{ph}$ as large as possible subject to this requirement
[61] W. Dieterich, A. Nitzan and K. Saito, unpublished.
[62] H. Haug and A.-P. Jauho, *Quantum Kinetics in Transport and Optics of Semiconductors* (Springer, Berlin, 1996).
[63] V. May and O. Kühn, *Charge and energy transfer dynamics in molecular systems* (Wiley-VCH, Berlin, 2000).
[64] A. Nitzan, *Chemical Dynamics in Condensed Phases* (Oxford Univ. Press, Oxford, 2006).
[65] C. Meier and D. J. Tannor, J. Chem. Phys. **111**, 3365 (1999).
[66] S. Welack, M. Schreiber, and U. Kleinekathofer, J. Chem. Phys. **124**, 044712 (2006).